\documentclass[12pt,preprint]{aastex}
\usepackage{amsmath}
\usepackage{graphicx}
\usepackage{epsfig}
\usepackage{color}
\usepackage{url}
\usepackage{indentfirst}
\usepackage{soul}
\begin{document}
\title{X-ray study of Variable Gamma-ray Pulsar PSR J2021+4026}
\author{Wang H.H.\altaffilmark{1}, Takata J.\altaffilmark{1},  Hu C.-P.\altaffilmark{2},
   Lin L.C.C.\altaffilmark{3}, Zhao J. \altaffilmark{1}}
\email{takata@hust.edu.cn}
\altaffiltext{1}{School of physics, Huazhong University of Science and Technology, Wuhan 430074, China}
\altaffiltext{2}{Department of physics, The University of Hong Kong, Pokfulam Road, Hong Kong}
\altaffiltext{3}{Department of Physics, UNIST, Ulsan 44919, Korea}
\begin{abstract}
  PSR~J2021+4026  showed a sudden decrease in the gamma-ray emission
  at the glitch that occurred around 2011, October 16, and a relaxation of the flux to the pre-glitch state at around 2014 December.  
  We report X-ray analysis results of the data observed
  by \emph{XMM-Newton} on 2015 December 20 in the post-relaxation state.
  To examine any change in the X-ray emission,
  we compare the properties of the pulse profiles and spectra
  at the low gamma-ray flux state and at the post-relaxation state. 
  The phase-averaged spectra for  both states can
  be well described by a power-law component plus a blackbody  component. 
  The former is dominated by unpulsed emission and is probably
  originated from the pulsar wind nebula as reported by Hui et al (2015).
  The emission property of the blackbody component is consistent with the emission from the polar cap heated by the back-flow bombardment of the high-energy electrons or positrons that were accelerated in the magnetosphere. 
  We found no significant change in the X-ray emission properties between two states. 
We suggest that the change of the X-ray luminosity is at an order of $\sim 4$\%, which is difficult to measure with the current observations. We model the observed X-ray light curve with the heated polar cap emission, and  we speculate that the observed large pulsed fraction is owing to asymmetric magnetospheric structure. 
  \end{abstract}

\section{Introduction}
A pulsar is a fast spinning and highly magnetized neutron star, which is a condensed star with an averaged mass density of $\sim 10^{14-15}{\rm g~cm^{-3}}$, and it is observable in radio to very high-energy TeV  gamma-ray bands. 
Although the radio emission is the main window to investigate the timing properties of the pulsars, all sky monitor of the Fermi Large Area Telescope (\emph{Fermi}-LAT, Atwood et al. 2009; Abdo et al. 2009a; Ackermann et al. 2012), which is a space observatory launched in 2008, enables us to perform a long term survey of the pulsars in the gamma-ray bands.  The \emph{Fermi}-LAT has observed the gamma-ray emission from $>200$ pulsars\footnote{https://confluence.slac.stanford.edu/display/GLAMCOG/Public+List+of+LAT-Detected+Gamma-Ray+Pulsars}\footnote{https://fermi.gsfc.nasa.gov/ssc/data/access/lat/fl8y/} (Abdo et al. 2013). In particular, \emph{Fermi}-LAT uncovered many new gamma-ray pulsars in the Cygnus region, and the most intriguing one among them is PSR J2021+4026.  
PSR J2021+4026 is an isolated pulsar that  belongs to the Geminga-like pulsars
(lacking radio-quiet emission with certain detections of pulsed detections in both the
X-ray and gamma-ray bands (Lin 2016). It  is also known as the first variable gamma-ray pulsar seen by the \emph{Fermi}-LAT.  
It is associated with the supernova remnant G78.2+2.1 (Abdo et al. 2009b). This pulsar has a spin period of $P=265$~ms and a spin-down rate of $\dot{P} = 5.48 \times10^{-14}$, corresponding to a characteristic age of $\tau_c \sim 77$~kyr, a surface dipole field of $B_{d}\sim 4\times10^{12}$~G, and a spin-down power of $\dot{E}\sim 10^{35}\,{\rm erg/s}$.

Allafort et al.(2013) reported results of a detailed analysis on the gamma-ray emission from PSR J2021+4026, and they found a glitch around  MJD~55850 (2011 October 16) on a timescale shorter than one week. 
This glitch increased the spin-down rate from  $ |\dot{f}|=(7.8 \pm 0.1) \times 10^{-13} {\rm Hz~s^{-1}} $ to
$ (8.1 \pm 0.1) \times 10^{-13} {\rm Hz~s^{-1}}$ .  
Moreover, the glitch accompanied with (1) a decrease of  flux ($>$ 100MeV) 
   by $\sim 18 \%$,  from $(8.33 \pm 0.08) \times 10^{-10}  {\rm erg~cm^{-2}
     s^{-1}} $  to $ (6.86 \pm 0.13) \times 10^{-10} {\rm erg~cm^{-2}  s^{-1}}$,  (2) a significant change in the pulse profile $(> 5\sigma)$, and (3) a marginal change in the gamma-ray spectrum $(< 3\sigma)$. 
   Before the glitch, the pulse profile consisted of two strong peaks plus a small third peak in the bridge region.
   After the glitch, there was no evidence of the third peak. 
   Ng et al. (2016) reanalyzed the $\emph{Fermi}$-LAT data with a longer time span. They found that the flux drop
   caused by the glitch was a permanent-like effect, and the low gamma-ray flux state continued 3~years after  the glitch.  
   Zhao et al. (2017) reported the timing analysis of $\sim$8-year \emph{Fermi}-LAT data,
   and they found the relaxation at around 2014 December, where  the spin-down rate and the gamma-ray emission returned to the state before the glitch. The pulse profile and spectrum after the  relaxation are
   consistent with those before the glitch.

   An unidentified X-ray source 2XMMJ202131.0+402645 had been investigated as a promising counterpart of PSR J2021+4026 (Trepl et al. 2010; Weisskopf et al. 2011), and a deep observation was done by \emph{XMM-Newton} (ESA: XMM-Newton SOC\footnote{``XMM-Newton Users Handbook'', Issue 2.15, 2017}) at MJD 56,028, when
   the pulsar stayed at the low gamma-ray flux state after the glitch.
   The detected spin frequency  is consistent with the gamma-ray pulsation of PSR J2021+4026 at the same epoch, proving 2XMMJ202131.0+402645 is indeed the counterpart of PSR J2021+4026 (Lin et al. 2013). The  single broad pulse profile and
   the blackbody spectrum  with an  effective temperature of $kT_B\sim 0.25$ keV and
   radius $R_{eff}\sim 300$m imply that the X-ray  emission came from the polar cap heated
   by the back-flow bombardment of
   the high-energy electrons or positrons that were  accelerated in the magnetosphere.

    Emission from the heated polar cap will be closely related to the gamma-ray emission process. In the outer
   gap
   model, for example, the pair-creation process inside the accelerator produces the electrons and positrons, and
   the electric field along the magnetic field separates the pairs, and results in the formation of the in-going and out-going particles (Zhang \& Cheng 1997). The created pairs inside the gap are accelerated to a Lorentz factor of $>10^7$ and
   emit the gamma-rays via the curvature radiation process. Half of the created pairs will return to  and heat up the
   polar cap region. In this model, the flux of the observed X-ray  emission will be proportional to the gamma-ray flux.
   Hence, it is  expected   that the state change in the spin-down rate/gamma-ray emission is also accompanied with 
   the change in the observed X-ray emission.  In this paper,
   we report the results of a new X-ray observation performed by \emph{XMM-Newton} in the post-relaxation state,
   and compare the emission properties with the previous X-ray properties observed
   in the low gamma-ray flux state (section~\ref{analysis}).  In section~\ref{discuss},
     we will discuss  the X-ray light curve of the heated polar cap and
     the emission geometry.

\section{Data analysis}
\label{analysis}
\begin{figure}
  \centering
  \includegraphics{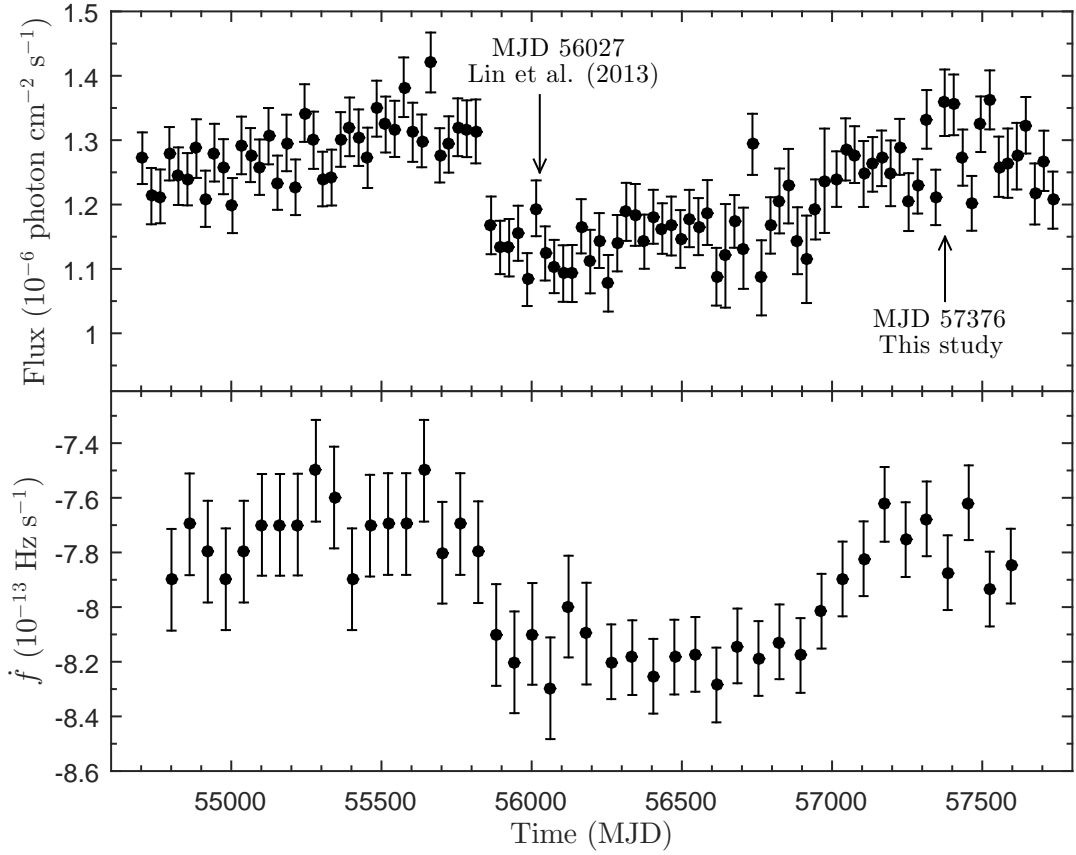}
  \caption{About an eight-year evolution of the gamma-ray flux (top) and the spin-down rate (bottom).
    The epochs of the two X-ray observations are indicted in the top panel. The figure is adopted from
     Zhao et al. (2017)}
  \label{flux_fdot}
\end{figure}
We analyze the archive data taken by \emph{XMM-Newton} on 2015 December 20 (MJD~57,376, Obs. ID: 0763850101, PI: Razzano),
which is  about 3.7 years after the previous  \emph{XMM-Newton} observation performed in
the low gamma-ray state (Lin et al. 2013), and  about 1 year after  the relaxation in 2014 December (Figure~\ref{flux_fdot}).
This new observation was performed with a total exposure of $\sim$~140ks.
The MOS1/2 CCDs were operated in the full-window mode (time resolution 2.6s), and PN CCD
was operated in the small-window mode (time resolution 5.7ms). Only PN data
enables the timing analysis of this pulsar. Event lists from the  data are produced  in the standard way using the  most
updated instrumental calibration and the  $emproc/epproc$ tasks of the  \emph{XMM-Newton} Science Analysis Software
(XMMSAS, version 16.0.0). After filtering the 
events,  which are potentially contaminated, the effective exposures are 134~ks and 130~ks for MOS1 and MOS2, 94~ks for PN,
respectively. 
A  point source is significantly detected ($>10\sigma$) by XMMSAS task $edetect\_chain$ at the position of PSR J2021+4026.
To perform the spectral and timing analyses, we extract EPIC data from circles with a radius of $20''$  centered at its
nominal X-ray position $({\rm R.A.}, {\rm decl.})=(20^{\rm h}21^{\rm m}30^{\rm s}.733, +40^{\circ}26'46''.04)$ (J2000). 

\subsection{Timing analysis}
   \label{timing}
 \begin{figure}
   \centering
   \includegraphics{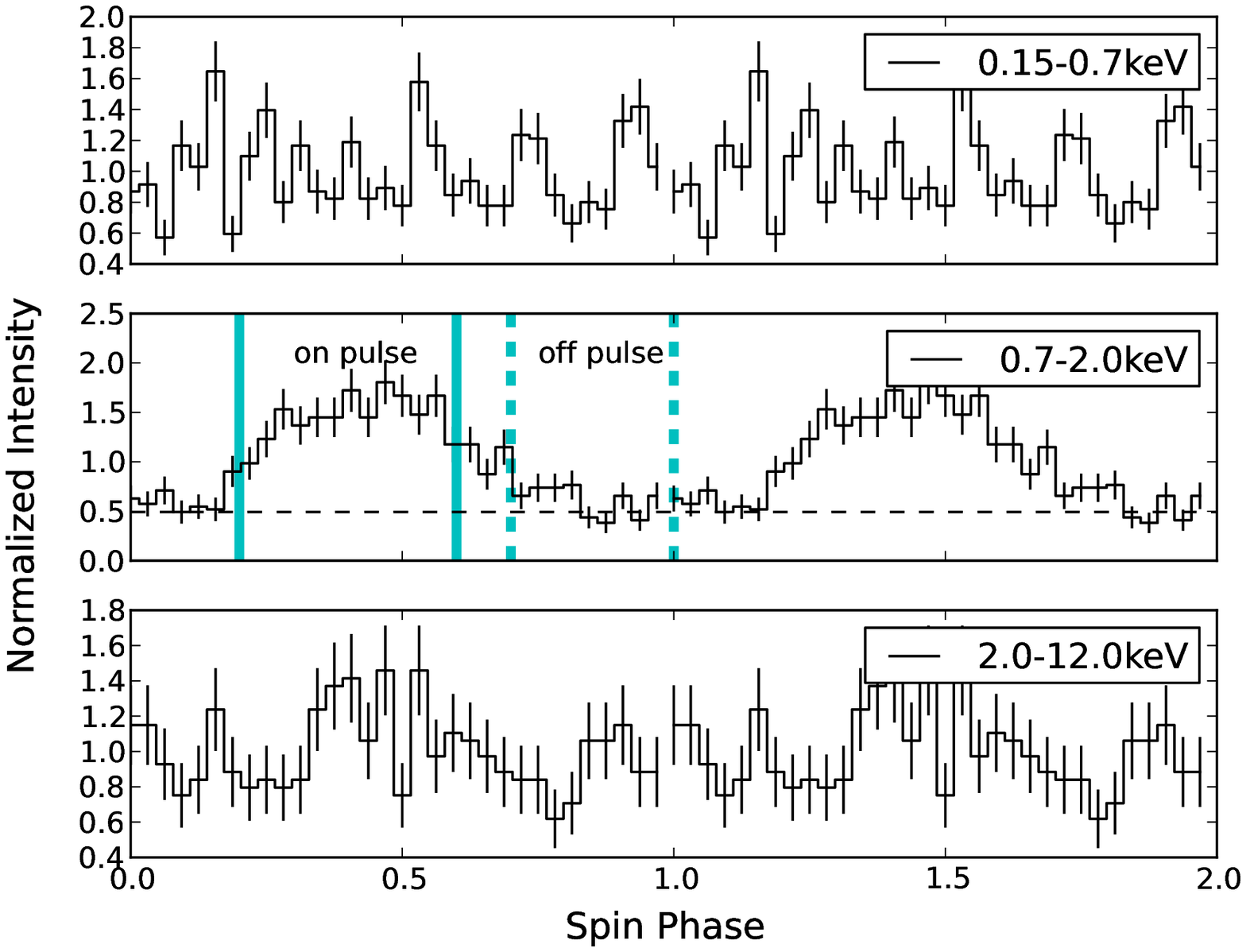}
   \caption{Energy dependent X-ray folded light curves of PSR~J2021+4026 in the energy bands
     0.15-0.7~keV (upper), 0.7-2.0~keV (middle) and 2.0-12~keV  (bottom).
     Two cycles are presented for clarity. In the middle panel, the vertical solid and dashed
       lines define the on-pulse and off-pulse phase, respectively, and the horizontal dashed line shows a background level
       determined by nearby source free region. We normalize the count so that the average intensity is unity, indicating the background emission occupies
   about 50\% of the   0.7-2.0keV emission from  the direction of the source.}
   \label{fig}
 \end{figure}

  \begin{figure}
     \centering
     \includegraphics[scale=0.6]{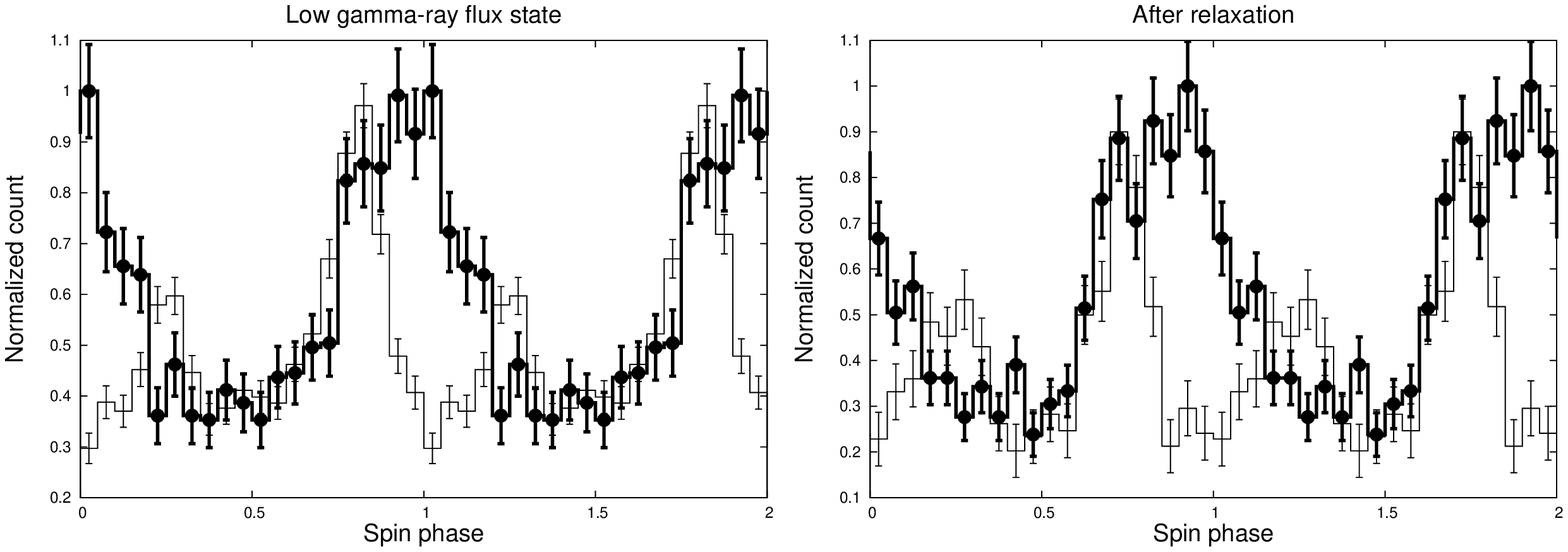}
     \caption{Folded light curves of PSR J2021+4026 in 0.7-2.0~keV (thick histograms) and $>$0.1~GeV (thin histograms) energy bands. Left panel: Low gamma-ray flux state. Right panel: post-relaxation state.}
       \label{lightgx}
       \end{figure}

  Our main purpose is to investigate the change in the pulsed X-ray emission before and after the relaxation occurred
  at around 2014 December. Therefore, a timing analysis is crucial. Following Lin et al. (2013), we divide the PN data into three energy bands; 0.15-0.7~keV, 0.7-2.0~keV and 2.0-12~keV. 
  For the timing analysis, the available photon number is 1399 counts for 0.15-0.7~keV,  1170 counts
  for 0.7-2.0~keV and 724 counts for 2.0-12~keV.
  The arrival times of all the selected events are barycentric-corrected with the aforementioned X-ray position
  and the latest DE405 Earth ephemeris.

  In the analysis,  the pulsation is significantly detected  only  in 0.7-2.0~keV energy bands, and
  no conclusive evidence of the pulsation is obtained in the energy bands of 0.15-0.7~keV and 2.0-12~keV.
   Based on the Rayleigh test (Mardia 1972; Gibson et al. 1982) applied for 0.7-2.0~keV observation,
  a significant peak is found at  $f=3.7688991(2)$Hz with $Z_1^2=196$, where we assessed the uncertainty with eq. 6(a) provided in Leahy (1987),  using total duration of $\sim$100~ks. 
  In 0.15-0.7~keV and 2.0-12~keV energy bands, the X-ray pulsation is insignificant with  $Z_1^2\sim 5$.
  This spin frequency  is consistent with that derived by the $Fermi$-LAT (Zhao et al. 2017).

  Figure~\ref{fig} shows the folded light curves in three energy bands; 0.15-0.7~keV (top), 0.7-2.0~keV (middle), and 2.0-12~keV (bottom).
  In the figure,   the folded light curve in 0.7-2.0~keV bands shows a broad and single peak pulse profile.
  In Figure 3, we compare the light curves  in 0.7-2.0~keV energy bands
  in the low gamma-ray flux state (right panel) and in the post-relaxation state (left panel). 
  Within the current quality of the data, no significant difference can be seen in the two pulse profiles.

    In the X-ray light curve of the 0.7-2.0~keV energy band (middle panel of Figure~\ref{fig}), we determine a background level (horizontal dashed line)
    with a nearby source-free region, and we can find that the background
    emission explains
    about 50\% of the emission from the direction of the source and it
    explains the emission at  the off-pulse phase. It is expected,
    therefore, that the pulse profile after subtracting
    the background has a large pulsed fraction, which is defined by
    $F=(f_{max}-f_{min})/(f_{max}+f_{min})\times 100$\% with
    $f_{max}$ and $f_{min}$ being the maximum and minimum count rates, respectively. 
    The large pulsed fraction ($F\sim 100$\%) can be used to constrain the emission geometry by
    modeling the light curve of the heated polar cap emission (section~\ref{discuss}).

    Lin et al. (2013) compared the pulse phases in X-ray and gamma-ray bands in the low gamma-ray flux state, and found that the X-ray peak lags the stronger gamma-ray peak (the second peak) and proceeds after
    the smaller peak (the first peak).
  To investigate the phase relation between the X-ray and gamma-ray pulses  in the post-relaxation state, we extract the $Fermi$-LAT data of about a half year (MJD~57,283--57,471) centered at the epoch of the X-ray observation, and create local ephemeris (Table 1).   
  With the standard process for the data reduction using the $Fermi$-Scinece tools v10r0p5 package (Zhao et al. 2017 for more details of PSR J2021+4026), we assign the probability of each photon originating from the pulsar. 
  After performing the $gtbary$ task for barycentric time corrections to photon arrival times,
  we convert the arrival time to the pulsar spin phase using the local ephemeris created by \emph{Fermi}-LAT data (i.e., Table~1).

  Figure~\ref{lightgx} summarizes the light curves of PSR J2021+4026 in the X-ray (thick histograms) and gamma-ray  (thin histograms) using the same ephemeris reported in Lin et al. (2013) for the low gamma-ray flux state (left panel) and the ephemeris in Table~1 for the post-relaxation state (right panel).
  As we can see in the figure, the phase correlation between the X-ray peak and the gamma-ray peak in the post-relaxation state
  is very similar to that in the low gamma-ray flux state, that is,  the X-ray peak lags the stronger
  gamma-ray peak (the second peak) and proceeds after the smaller peak (the first peak). The cross-correlation  coefficient
  attains the maximum value at a phase lag of $\sim -0.14$ for the both states.
  Hence, no significant change in the correlation of the X-ray and  gamma-ray pulsations before and after the relaxation is found.

\begin{table}[ht]
  \centering
\begin{tabular}{cr}
  \hline  \hline
  Parameters           \\
  \hline
  Right ascension       & $20^{\rm h}21^{\rm m}30^{\rm s}.733$ \\
  Declination    & $+40^{\circ}26'46''.04$\\
  Valid MJD range       & 57,283$\sim$57,471\\
  Pulse frequency, $f ({\rm Hz})$   & 3.7688994400(7)  \\
  First derivative of pulse frequency, $\dot{f}({\rm s^{-2}})$
  &-7.707(4)$\times 10^{-13}$ \\
  Second derivative of pulsar frequency, $\ddot{f}$ (s$^{-3}$)
  & $6.0(9)\times10^{-22}$ \\
  Third derivative of pulsar frequency, $\dddot{f}$ (s$^{-4}$)
  & $2(9)\times10^{-29}$ \\
  Epoch zero of the timing solution (MJD)  & 57,377 \\
  RMS timing residual ($\mu$s)  & 1677.862\\
  \hline
\end{tabular}\\
  \caption{Ephemeris of PSR J2021+4026.}
   \end{table}
 \subsection{Spectral analysis}

 \begin{table}
   \centering
   \begin{tabular}{crr}
     \hline \hline
     Parameters   & MJD~56,028    & MJD~57,376\\  \hline
     ${\rm N_{H}(10^{22}cm^{-2})} $   & \multicolumn{2}{c}{$0.9_{-0.3}^{+0.4}$} \\
     Flux$_{bb} (10^{-13}{\rm erg~cm^{-2}s^{-1}})$\tablenotemark{a}& $1.4_{-0.6}^{+1.4}$ & $1.6_{-0.7}^{+1.5} $\\
     Flux$_{p}  (10^{-14}{\rm erg~cm^{-2}s^{-1}})$\tablenotemark{b}  & $ 3.5_{-0.7}^{+1.3}$   & $4.7_{-1.0}^{+1.4}$ \\
     $kT_B$ (kev)     & $0.21 _{-0.03}^{+0.03}$     & $0.21_{-0.04}^{+0.04}$  \\
     $R_{eff}$ (m)\tablenotemark{c}   & $272_{-255}^{+591}$  & $255^{+622}_ {-245}$   \\
     Photon index  & $1.0^{+0.7}_{-0.8}$      & $1.3_{-0.9}^{+0.9}$ \\
     $\chi ^{2}/$dof  & \multicolumn{2}{c}{105.89/114}\\
     \hline
         {\footnotesize a: Flux of blackbody component in 0.2-12~keV ~~~~~~~~~~~~~} &&\\
         {\footnotesize b: Flux of power law component in 0.2-12~keV ~~~~~~~~~~~~~~} &&\\
           {\footnotesize c: Effective radius at $d=1.5$~kpc~~~~~~~~~~~~~~~~~~~~~~~~~~~~~~~~} &&\\
   \end{tabular}
   \label{tab:booktabs}
   \caption{Parameters of the phase-averaged spectra determined in the low gamma-ray flux state (middle column) and in the post-relaxation state (third column). The uncertainties of each spectral parameters are assessed in 1$\sigma$ for four parameters of interest for the multi-component model.}
     \end{table}

\begin{table}
 \centering
 \begin{tabular}{ crr }
   \hline \hline
   parameter   & MJD~56,028    & MJD~57,376\\  \hline
   Flux ($10^{13}{\rm erg~cm^{-2}s^{-1}}$)\tablenotemark{a} &  $1.3_{-0.3}^{0.3}$ & 
   $1.0_{-0.3}^{+0.4} $\\
   $kT_B$ (keV) & $0.27_{-0.03}^{+0.04} $      & $0.27_{-0.04}^{+0.05}$ \\
   $R_{eff}$ ($m$)$\tablenotemark{b}$ & $234_{-168}^{+240}$   & $214_{-163}^{+258} $ \\
   $\chi ^{2}/$D.O.F   & \multicolumn{2}{c}{17.13/17}\\
   \hline
       {\footnotesize a: Flux in 0.2-12~keV} ~~~~~~~~~~~~ &&\\
       {\footnotesize b: Effective radius at $d=1.5$~kpc} &&\\
 \end{tabular}
 \label{tab:booktabs}
 \caption{Parameters of the pulsed spectra in the low gamma-ray flux state (middle column) and
   in the post-relaxation state (third column). The hydrogen column density is fixed at $N_H=7\times 10^{21}~{\rm cm^{-2}}$.}
\end{table}

  \subsubsection{Phase-averaged  spectrum}
  In order to further investigate the X-ray emission properties from this pulsar,
  we carry out the spectral analysis of all three EPIC cameras (MOS1/2 and PN),
  and we compare with the previous results given in the low gamma-ray flux state.
  We generate the spectra of MOS1/2 and PN from photons in the 0.15-12keV energy bands
  within a radius of 20$''$ circle centered at the source.
  The background spectra are generated from a nearby region of  the same size of  the corresponding CCD. 
  The response files are generated by the XMMSAS tasks $rmfgen$ and $arfgen$.
  We group the channels so as to achieve the signal-to-noise ratio S/N=3 in each energy bin.

\begin{figure}
      \centering
      \epsscale{1.0}
      \begin{tabular}{@{}cc@{}}
        \rotatebox{-90}{\includegraphics[width=0.6\textwidth]{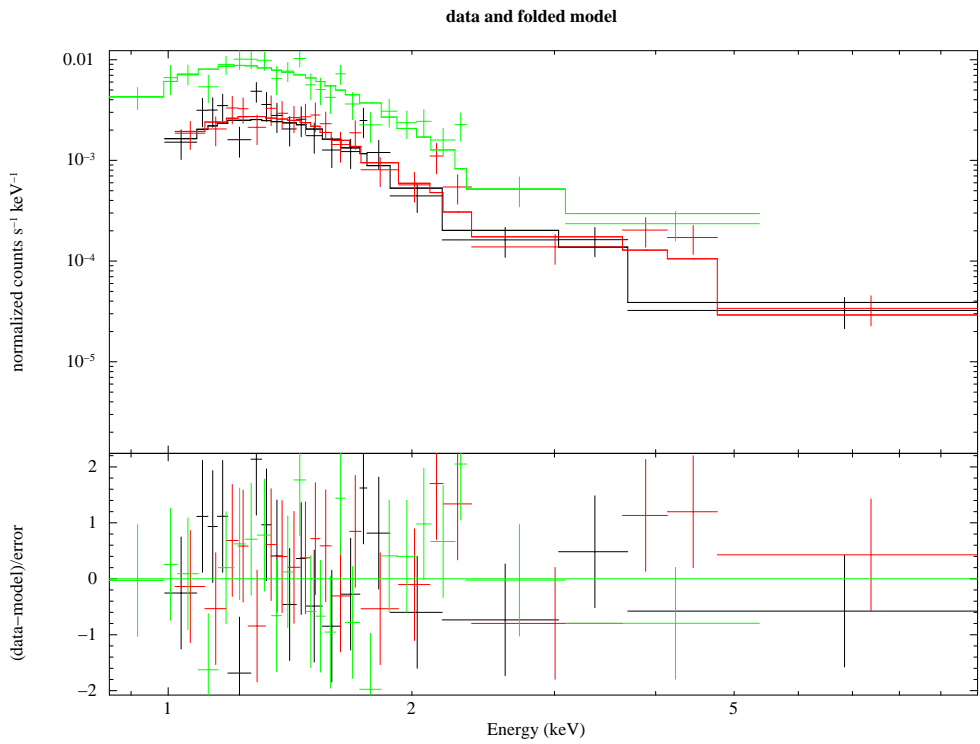}} 
      \end{tabular}
      \caption{Phase-averaged of spectrum of PSR~J2021+4026 in 0.5-10~keV. The observed spectra with the PN (green spectrum) and MOS1/2 detectors (black and red spectra) are simultaneously fitted by an absorbed blackbody plus power-law model.} 
      \label{spectrum}
    \end{figure}

  Figure~\ref{spectrum} shows the phase-averaged spectrum in the post-relaxation state taken by all three cameras.
To investigate any changes of the spectral properties in the low gamma-ray flux state and the post-relaxation state, we fit the data in the two states simultaneously using the XSPEC (version 12.9.1). Since the interstellar hydrogen is the main absorber of X-rays, we expect that the hydrogen column density ($N_{{\rm H}}$)
is not changed by the state change. In the analysis, therefore,  we tie the column densities  in the two data sets, and leave all other parameters as free.  First, we fit the data with the single-component model, that is, a single power=law (PL) or a single blackbody (BB) radiation model.
  We found that the single-component model cannot provide appropriate fits; for the power-law model, the best-fit power law index is very large ($\Gamma\sim 4$) with $\chi^{2}=162$ for 121 degrees of freedom (dof), which is far steeper than that ($\Gamma\sim 1.5$) predicted  by the synchrotron radiation from the secondary electron/positron pairs (Cheng \& Zhang 1999). 
  The blackbody model also  provides an unacceptable fit with $\chi^{2}=205$ for 121 dof.

\begin{figure}
   \includegraphics[scale=0.6,angle=270]{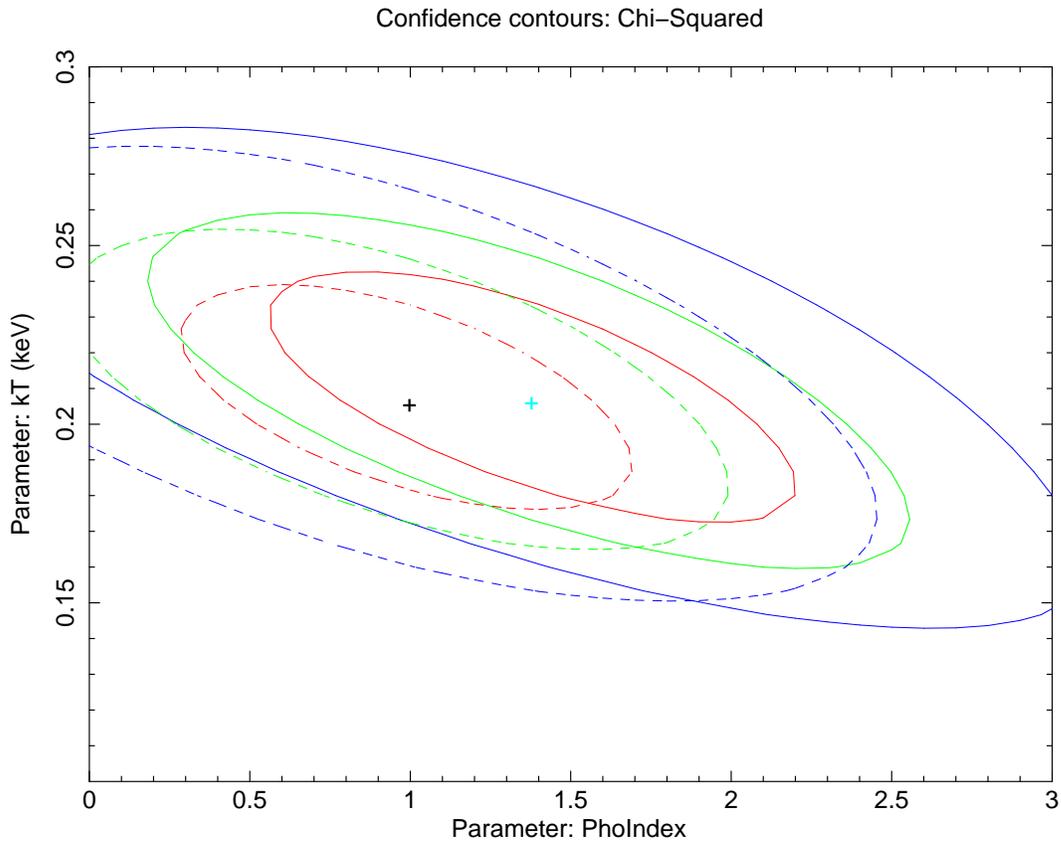} 
   \caption{$1\sigma$, $2\sigma$ and $3\sigma $ error contours of the spectral parameters
     for the low gamma-ray flux  state (dash line) and post-relaxation state (solid line).
     The black and blue cross symbols indicate the parameters providing the minimum $\chi^2$ for the low
   gamma-ray flux state and post-relaxation state, respectively.}
   \label{contour}
     \end{figure}

  Since Lin et al. (2013) obtained a reasonable fit with the PL+BB model, we also apply the PL+BB model for the simultaneous fitting.  We found that  the PL+BB model can describe the data in the low gamma-ray flux state and the post-relaxation state simultaneously, and
  the best fitting parameters of the X-ray data in both states are consistent with  each other within 1$\sigma$ error (Table~2); the observed spectra are well fitted with the photon index $\Gamma\sim 1-1.3$, the surface temperature of $kT_B\sim 0.25-0.3$keV and
  the effective radius of $R_{eff}\sim 250-300$m. The observed flux in the energy band of 0.2-12~keV is dominated by the blackbody component.   The result is consistent with that indicated in Lin et al. (2013), and the power-law component is probably came from
  the pulsar wind nebula (Hui et al. 2015).  Figure~\ref{contour} shows
  the confidence contours  on the plane determined by the photon index and the temperature. 
  We can see in the figure that the distributions of the errors are  very similar as well, suggesting no significant evolution of the spectral properties.

     \subsection{Spectrum of the pulsed component}
     \label{pulsed}
\begin{figure}[ht]
    \centering
    \includegraphics[scale=0.6,angle=270]{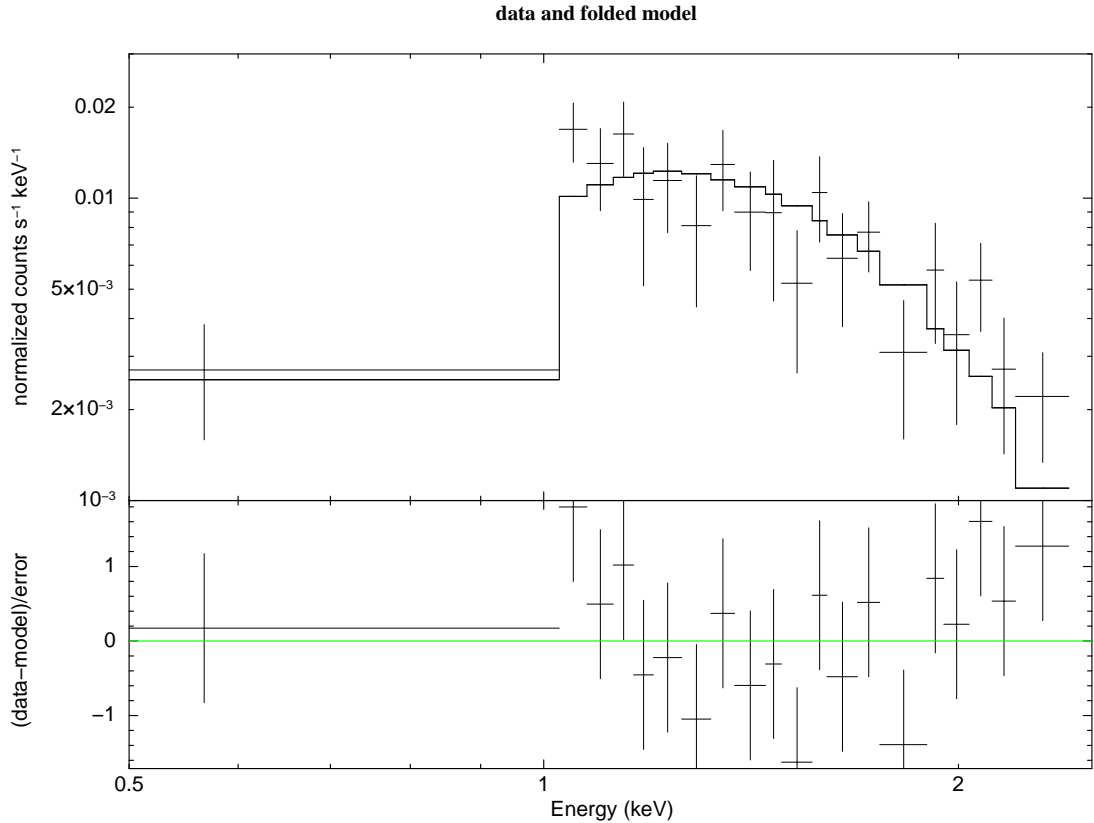} 
    \caption{Pulsed spectrum of PSR J2021+4026 in the post-relaxation state. This spectrum is obtained by subtracting the data of off-pulse phase from the on-pulse phase and provides the best-fit for a single blackbody model in the energy band of 0.2-7 keV. The lower panel demonstrates the $\chi ^2$ fit statistic.}
\label{extract}
 \end{figure}

   We analyze the data in the low gamma-ray flux state and in the post-relaxation state, and
    we  compare the spectra of the pulsed component for two states. 
   According to Figure~\ref{fig} of the post-relaxation state, we define the phase intervals of 0.2-0.6 and 0.7-1.0 as the ``on-pulse''
   and ``off-pulse'' (i.e., DC level) components, respectively.
   We also use the same intervals in the light curves of the low gamma-ray flux state. 
   The event files in each phase are extracted using $xronwin$ and $xselect$, and the spectrum of the pulsed component is obtained by subtracting
   the spectrum of ``off-pulse'' phase from that of the ``on-pulse'' phase. 
   Because of the limited photon counts, we grouped the channels  so as to archive the signal-to-noise ratio S/N=2
   in each energy bin. The  generated spectra (Figure~\ref{extract}) are  fitted by the single  BB model, and its
   best-fit parameters (Table~3) are $kT_B\sim 0.26$~keV and $R_{eff}\sim 250$~m at $d=1.5$~kpc.  The result in low gamma-ray flux
   state is fully consistent with the result in Hui et al. (2015). The single power-law model
   also provides an acceptable fit. However, the best-fit photon index is very large ($\Gamma\sim 3.8$)
   with $\chi^2=8.41$ for  15 dof, again which is far steeper than the theoretical prediction ($\Gamma\sim 1.5$).
   The two component (BB+PL)  model fits the data with $kT_B\sim 0.19$keV, but the model cannot constrain the photon index.
   Based on our fitting presented in Table~3, no significant change is
   found in the pulsed spectra before and after relaxation of the state.

 \section{Discussion and summary}
\label{discuss}
 \subsection{Emission from the heated polar cap}
 We have compared the properties of the  X-ray emission observed at MJD~56,028 in the low gamma-ray flux state and at MJD~57,376
 observed in the post-relaxation state.  Although both spin-down rate
 and  gamma-ray flux observed at two epochs are different, our measurements find
 no significant change in any of the X-ray emission properties.
  The blackbody radiations in both states are  measured with an effective radius $R_{eff}\sim 300$~m at  $d=1.5$~kpc, which is comparable to the theoretical prediction of the polar cap size, $R_{pc}\sim R_{NS}(R_{NS}/R_{lc})^{1/2}\sim 280$~m,  where
  $R_{NS}=10^{6}$~cm is the neutron star radius, and $R_{lc}=cP/2\pi$ is the light cylinder radius that can be derived from the spin period ($P$) and the light speed ($c$).
  As proposed in the previous studies (Lin et al. 2013; Hui et al. 2015), the observed X-ray emission is likely originated from the polar cap heated by the bombardment of the high-energy electrons or positrons that were  accelerated in
  an acceleration region.

  $Fermi$-LAT found that the gamma-ray flux above the cutoff energy at around $\sim$3GeV decays slower than a pure exponential function (Abdo et al. 2013). This cutoff behavior favors the emissions from the outer magnetosphere
  [e.g. Arons (1983) for the slot gap model, Cheng et al. (1986ab) for the outer gap model, and  Spitkovsky (2006) for
    the current sheet], and it rules out the classical polar cap scenario (e.g. Daugherty and Harding 1996),
    which predicted a super exponential cutoff feature in the GeV spectrum because of the magnetic pair-creation process.
    In the outer gap model (Cheng et al. 1986ab; Zhang \& Cheng 1997; Takata et al. 2010), for example, 
     the luminosity of the gamma-ray from the accelerated electrons/positrons and  of the
     heated polar cap are characterized by the so-called fractional gap width, $f_{gap}$, which is defined by
     the ratio of the angular size of the gap thickness measured on the stellar surface to the angular
     size of the polar cap (i.e., $\theta_p=\sqrt{R_{NS}/R_{lc}}$).  
     The electrodynamics of the outer gap expects that  the radiation power of the gamma-ray emission is of the order of
     \begin{equation}
       L_{\gamma}\sim I_{gap}\times V_{gap}\sim f^3_{gap}L_{sd},
       \end{equation}
where $I_{gap}$ and $V_{gap}$ are the electric current and potential drop along the magnetic field line, respectively. 
In the outer gap, they may be of the  order of
$I_{gap}\sim \pi f_{gap} B_{d}R_{NS}^3/(P R_{lc})$ and $V_{gap}=f^2_{gap}
B_dR_{NS}^3/(2R_{lc}^2)$, respectively, where $B_d$ denotes the surface magnetic dipole field.
The gamma-ray emission  from PSR~J2021+4026 ($L_{sd}\sim
1.2\times 10^{35}\,{\rm erg~s^{-1}}$) is measured with a luminosity of 
$L_{\gamma}\sim 5\times 10^{34}(d/1.5{\rm kpc})^2(\Delta\Omega/3{\rm radian})\,{\rm erg~s^{-1}}$, where
$\Delta\Omega$ is the solid angle of the gamma-ray beam. 
The fraction gap thickness, therefore, is estimated as $f_{gap}\sim 0.75$, indicating that a large fraction of the spin-down power is converted into the high-energy radiation. 

The electrons and positrons in the outer gap  are  separated by the electric field parallel to the magnetic field line, and an half of the accelerated particles will be returned to the polar cap.
The number of  particles returned to the polar cap is  of the order of
\begin{equation}
  \dot{N}_e\sim \frac{I_{gap}}{2e}\sim7\times 10^{29}f_{gap} P^{-2}B_{12}\,{\rm s^{-1}},
\end{equation}
where $B_{12}$ is the surface magnetic field in units of the $10^{12}$G. Because of the curvature radiation process during the travel from the inner boundary of the outer gap to the  stellar surface, each return particle carries
only $10.6P^{1/3}$ ergs onto the stellar surface (Halpern \& Ruderman 1993; Zhang \& Cheng 1997). As a result, the luminosity of the X-rays from the heated
polar cap region is estimated by
\begin{equation}
L_{X} \sim 10.6P^{1/3}{\rm erg}\cdot \dot{N}_e\sim  10^{31} f_{gap}B_{12}P^{-5/3}\,{\rm ergs~s^{-1}},
\end{equation}
which gives  the order of $L_X\sim 10^{32}\,{\rm erg~s^{-1}}$ for PSR~J2021+4026.
Since the X-ray emission from PSR~J2021+4026 is likely originated from the polar cap heated by
the back-flow particles that were accelerated in the outer magnetosphere, it is expected
that the state change in the spin-down rate/gamma-ray emission  accompanies with 
the change in the heated polar cap  emission. In the low gamma-ray flux state, the spin-down rate was
$\sim 4\%$ higher  than one in the  pre-glitch and post-relaxation states.  We speculate that the change in the
magnitude of the global current caused the change in the spin-down rate at the glitch and at the relaxation,
and that the current in
the low gamma-ray flux state increased  by $\delta I_c/I_c\sim 4\%$. If we estimate the fractional gap thickness
from $f_{gap}=(L_{\gamma}/L_{sd})^{1/3}$, the observation implies that the fractional gap thickness decreased
by $\delta f_{gap}/f_{gap}\sim -8\%$. Since the X-ray luminosity is proportional to $f_{gap}I_c$, the expected change
in $L_{X}$ is found to be $|\delta L_{X}/L_{X}|\sim |\delta I_c/I_c+\delta f_{gap}/f_{gap}|\sim 4\%$, which is
smaller than the current uncertainty of our measurement ($>10$\%). Hence
it is  difficult to measure the change of the X-ray emission before and after the relaxation
with the current observations.

\subsection{X-ray/gamma-ray light curve model}
\begin{figure}
  \centering
  \includegraphics[scale=0.6]{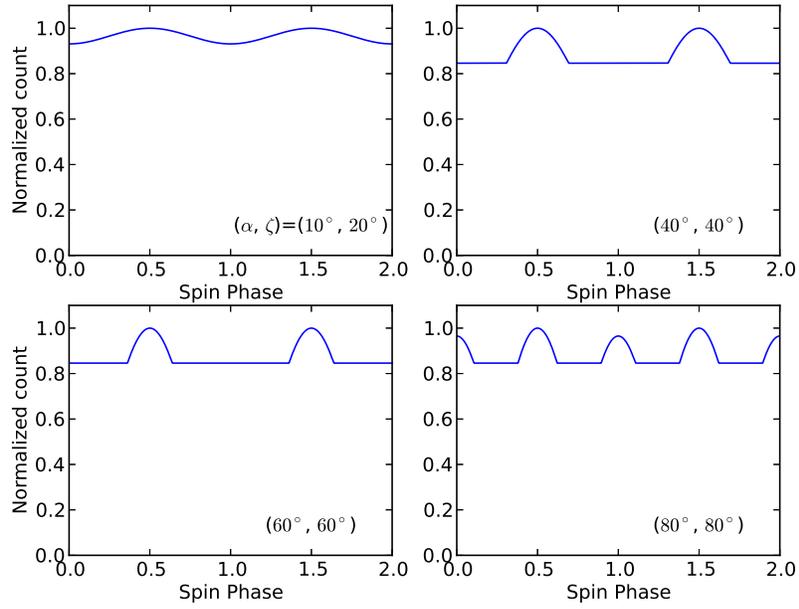}
  \caption{Model light curves for the  isotropic emission ($I(\theta)$=constant) from both of the  magnetic poles  heated by the return current. The
    assumed magnetic inclination angle $\alpha$ and the viewing angle $\zeta$ are indicated by the values in
  each panel. The light bending effect due to the neutron star gravity is taken into account with $R_{NS}/R_s=2.35$ (Beloborodov 2002; Bogdanov 2016). }
    \label{two-pole}
\end{figure}

\begin{figure}
  \centering
  \includegraphics[scale=0.6]{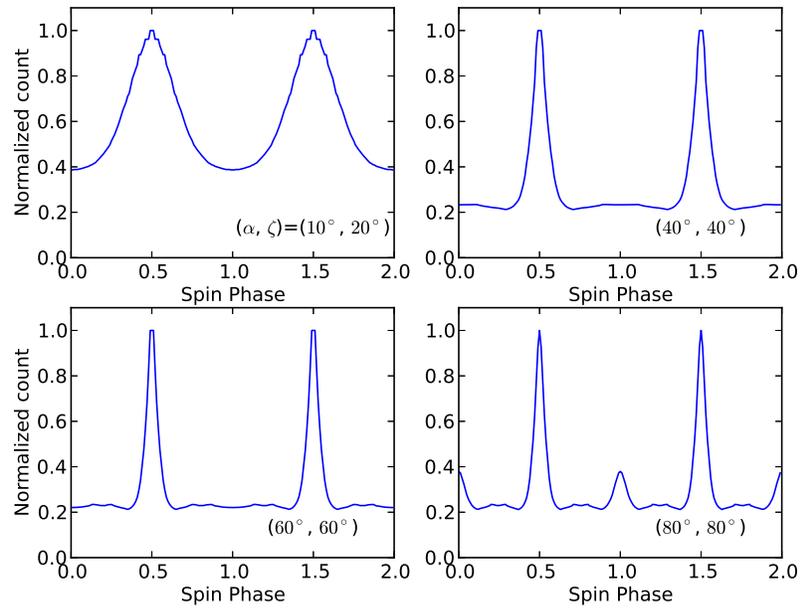}
  \caption{Same as Figure~\ref{two-pole}, but the local intensity depends
    on local emission angle.  The angular distribution
    is read from those of the models 2\&3 in Figure~2(b) of Zavlin et al. (1995).
  }
  \label{two-pole-1}
  \end{figure}

\begin{figure}
  \centering
  \includegraphics[scale=0.6]{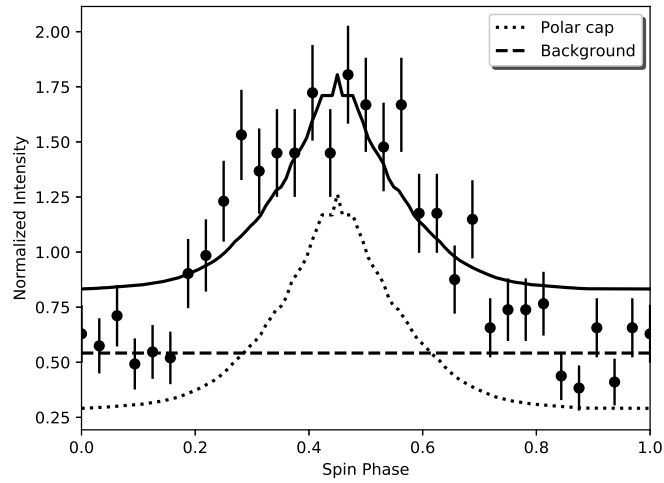}
  \caption{Comparing between the model and observed (0.7-2keV) light curves. For the heated polar cap emission (dotted line),
    the angular distribution of the intensity is approximated by the Figure 2(b)  of Zavlin et al. (1995).
    The background level (dashed line) is determined by the nearby source-free region. The model result is for $\alpha\sim 16^{\circ}$ and $\zeta\sim 20^{\circ}$, which is determined by the minimum chi-square. }
  \label{xfits}
\end{figure}

\begin{figure}
  \centering
  \includegraphics[scale=0.6]{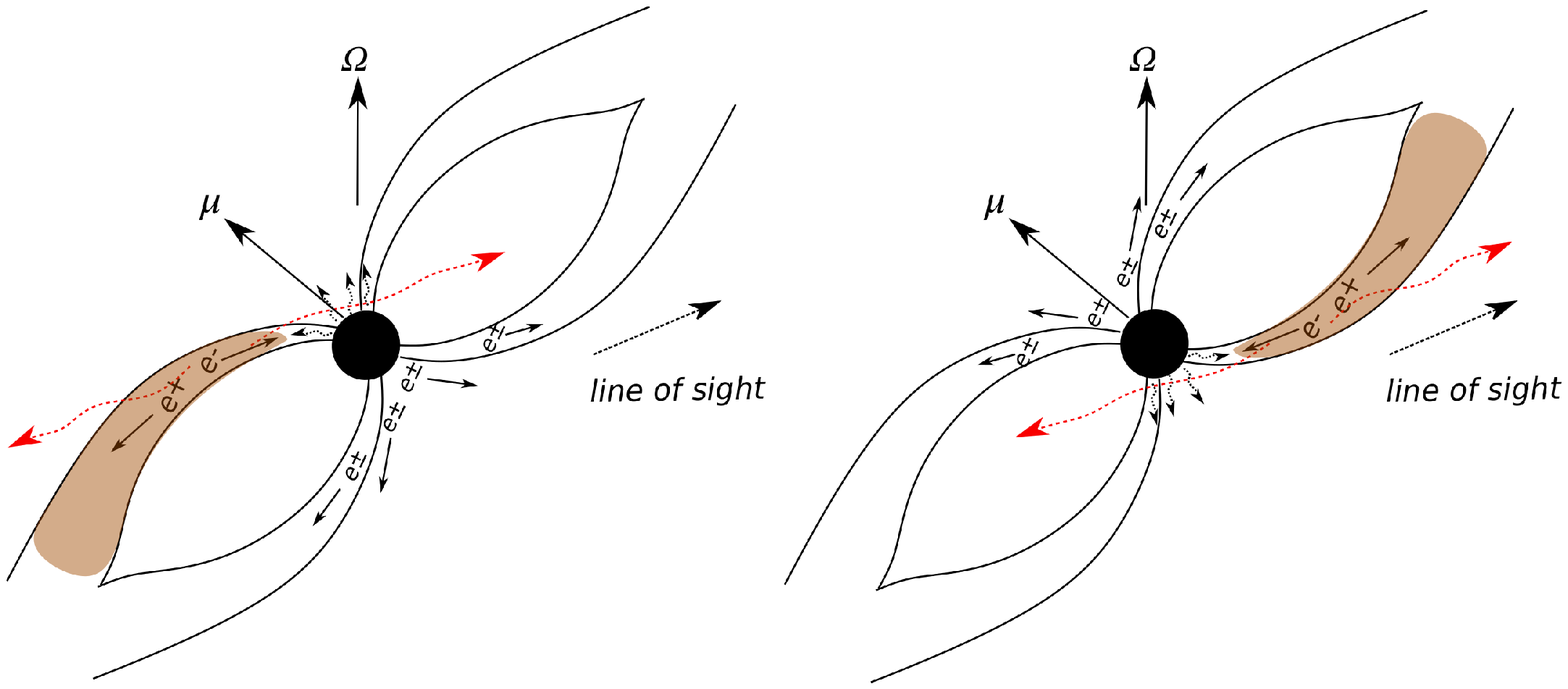}
  \caption{Possible geometry for PSR~J2021+4026. The charge separated pairs in the outer gap
    produce inwardly and outwardly propagating gamma-rays (red arrows), and the return particles heat up
    the polar cap region.  The X-ray pulse profile indicates that only
    single pole is heated up by return current from the outer gap, and  the outer gap in other
    hemisphere is quenched by the copious pairs from the polar cap region. The figure
    assumes $\zeta<90^{\circ}$ for the observer viewing angle,
    and only the  outer gap connecting to the north pole (left panel) or
    south pole (right panel) is active.    The double peak structure of
    the GeV light curve indicates the line  of sight is nearly 90 degree. }
    \label{geometry}
\end{figure}

As we have discussed above, the X-ray emission is likely originated from the heated polar cap region, and the GeV gamma-ray may be produced in the outer magnetosphere.  
The X-ray peak leading the first peak in the gamma-ray pulse profile will also
be  consistent with  the GeV gamma-ray
emission from the outer magnetosphere  (e.g. Romani \& Yadigaroglu 1995). 
The gamma-ray emission from PSR~J2021+4026 is observed with a double peak in the light curve separated by about a half spin phase.
Trepl et al. (2010) interpreted  the phase separation with the outer gap accelerator model and expected the viewing angle of $\sim 90^{\circ}$ measured from the rotation axis. 
Hui et al. (2015) assumed an orthogonal rotator with a single pole contribution to fit the X-ray light curve. 
Since the previous studies did not discuss in detail about the phase relation between the gamma-ray pulse and
X-ray pulse for PSR~J2021+4026, we compare the observed  X-ray/GeV light curves with the theoretical model.

Since the radius of the polar cap
region of PSR~J2021+4026 is of the order of 200-300m, which is much smaller than the neutron star size, we can safely ignore the size of the polar cap region, and assume the point-like hotspots at the magnetic poles. The theory including gravitational  light bending for the hotspot emission
on the neutron star surface has been developed in the previous studies (Pechenick et al. 1983;
Beloborodov 2002; Bogdanov et al. 2007; Bogdanov 2016).
In this paper, we apply an approximation relation (Beloborodov 2002; Bogdanov 2016), 
\begin{equation}
  \cos\psi\sim\frac{\cos\theta-R_s/R_{NS}}{1-R_s/R_{NS}},
\end{equation}
where $\psi$ and $\theta$ are the escape direction and the local angle of the emission direction, respectively. In addition
$R_s=2GM_{NS}/c^2$, and $M_{NS}$ is the mass of the neutron star. Such a formula can be used for $R_{NS}/R_{S}>2$.  In this paper, we
present a result with $R_{NS}/R_{s}$ is $\sim$ 2.35 for  typical $M_{NS}=1.4M_{\odot}$ and $R_{NS}=10$~km. The escape angle $\psi$ is related
to the spin phase as
\[
\cos\psi(t)=\sin\alpha\sin\zeta\cos\left(2\pi \frac{t}{P}\right)+\cos\alpha\cos\zeta
\]
where $\alpha$ and $\zeta$ are the angles of the magnetic axis and the line of sight measured from the spin axis.

We assume that the observed count at each spin phase in the light curve is proportional to $I_0(\theta)\cos\theta$, where $I_0(\theta)$ is the
local intensity. For the isotropic emission, $I_0(\theta)$=constant,  the previous studies (e.g. Beloborodov et al. 2002;  Bogdanov 2016) demonstrated
that the emissions from the two PCs, which are equally heated up, make the pulse profile with a small pulsed fraction  $F\sim 10\%$
(see also Figure~\ref{two-pole}). This is because
the light bending effect allows us to measure the emission from the backside of a neutron star and because
the observer can  almost always can see the emission from  two poles. A neutron star atmosphere model with a strong magnetic field
has been developed by the previous studies (e.g. Pavlov et al. 1994; Zavlin et al. 1995; Ho and Lai 2001, 2003;
Zane and Turolla  2006).
It is  discussed that the neutron star atmosphere
with a strong magnetic field makes the emission highly anisotropic, and the higher intensity emerges along the magnetic field. Zavlin et al. (1995) demonstrate that  the atmosphere with
a strong magnetic field can generate the light curve with a larger pulsed fraction and a shaper pulse width.

To compare the atmosphere-model light curve with the observation of PSR~J2021+4026,
we refer to the results of Zavlin et al. (1995), who
calculate the angular distribution, $I(\theta)$,  with the pulsar
parameters of $k_BT_{eff}\sim 0.25$keV and $B_d\sim 1.2\times 10^{12}$G.
Since their temperature of the heated polar cap is similar to that of PSR~J2021+4026 ($kT_{eff}\sim 0.21$keV), we may expect that
their results can represent  the case of  PSR~J2021+4026. Pavlov et al. (1994) compare between the calculated
angular distributions of the intensity  for  the nonmagnetic case  $B_d=0$ and the magnetic case $B_d=1.18\times 10^{12}$G. 
From their results,  we may expect that a slight
difference in the dipole magnetic
field ($B_d\sim 4\times 10^{12}$G for PSR~J2021+4026
comparing with  $B_s\sim 1.2\times 10^{12}$G of Zavlin et al. (1995))
does not significantly  change the angular distribution. Zavlin et al. (1995) present  the angular distributions for the photon energy $E_0=0.18$keV, 1.12keV and 2.47keV from Figure~2(b) in Zavlin et al. (1995). Since the significant  pulsation of PSR~J2021+4026
is detected in 0.7-2keV energy bands, we read the angular distribution of $E=1.12$keV from Figure~2 (b)
in Zavlin et al. (1995).  Finally, Zavlin et al. (1995) also show that the
angular distribution is less dependent on the neutron star model, that is,
$R_{NS}/R_s$. In this paper, therefore, we
refer their result of $R_{NS}/R_s=2.418$ (model 2 in Zavlin et al. 1995).
Figure~\ref{two-pole-1} summarizes the model light curves for the different set of the inclination angle and viewing angle.
By comparing  Figure~\ref{two-pole} with ~\ref{two-pole-1}, we  can see that the atmosphere with the magnetic field makes the light curve with a larger pulsed fraction ($F\sim 65$\%)
and a shaper pulse width, as demonstrated in the previous studies  (Pavlov et al. 1994; Zavlin et al. 1995). 

Figure~\ref{xfits} compares the model light curve and the
observed 0.7-1.2keV light curve; in the model, we
assume that 50\% of the total emission (solid line) is contributed
by the background (dashed line), as indicated by
the observed 0.7-2.0keV emission  (Figure~\ref{fig}).
We determine the inclination angle $\alpha\sim 16^{\circ}$
and the viewing angle $\zeta\sin 20^{\circ}$
to minimize the chi-square with the central points of the data.
Within the current simple treatment of the angular distribution, we find that although the size of the observational error  is relatively large, the pulsed fraction predicted in the model
light curve is smaller than that of the observation, as Figure~\ref{xfits} indicates. The pulsed fraction  indeed depends on many unknown parameters (e.g.  hotspot size, viewing geometry,
beaming effect, and $R_{NS}/R_s$). For example, we can expect that
the pulsed fraction decreases with the increase of the hotspot size and with the decrease of the
ratio $R_{NS}/R_s$. The pulsed fraction tends to increase
with the increase of the ratio $R_{NS}/R_s$.
The  observed pulsed fraction ($F\sim 100$\%), however, can be explained by unrealistically large
ratio $R_{NS}/R_s>5$ with the current beaming effect. With a reasonable ratio $R_{NS}/R_s$, a larger beaming effect will be required to give rise to the observed pulsed fraction.
A fine tuning of the parameters is
required to explain the observed X-ray light curve if two poles are equally heated up.

To obtain a more robust conclusion, it will be required to 
calculate the light curve integrated by the photon energy and with the magnetic field strength of PSR~J2021+4026, which will be done in the subsequent studies. We may argue, on the other hand,
that if both two poles are equally heated up by the return currents, the resultant
pulsed fraction would be smaller  than that observed for PSR J2021+4026, as Figure~\ref{xfits} indicates. 
Since it would  not be necessary that the magnetosphere is symmetric, the north and south
poles could have different temperatures, and the resultant light curve has  a large pulsed fraction.  We speculate that  the
asymmetry could be introduced by asymmetry of the magnetic field structure around the polar caps. It is argued that near the stellar surface,
the magnetic field configuration is not dominated by a dipole field (Ruderman \& Cheng 1988; Ruderman 1991). Higher order multipole field configuration is likely, and the strength of the multipole field  can be 1$\sim$3 orders of magnitude larger  than the global dipole field. This
could also affect the structure and magnetic pair-creation process
of the polar cap accelerator. For example, Timokhin and Harding (2015) point out that the multiplicity of the pair-creation cascade at the polar cap
accelerator is sensitive to the curvature and strength of the magnetic field. Moreover,
Harding \& Muslimov (2011a,2011b) argue that even a small distortion of the dipole field  and/or the offset of the polar cap from the dipole axis can greatly enhance the accelerating electric field and the resultant multiplicity of the pairs. It is
expected therefore that  an asymmetry of
the magnetic field of the two polar caps causes the asymmetry of
the polar cap accelerators.
If the polar cap accelerator supplies the return current
to heat up the polar cap, the temperatures of the two polar caps
could be different. 

The return particles can also be supplied by the accelerator around the light cylinder, which is probably the emission site of
the observed GeV gamma-rays. For example, the outer gap accelerator is a possible region to supply the returning particles.
It has been argued that the outer gap structure is sensitively controlled by the electrons or positrons that enter into the gap from
the outside along the magnetic field lines (Hirotani \& Shibata 2001; Takata et al. 2016). Moreover, the outer gap
will be quenched  (or less luminous),  if the particles with a super Goldreich-Julian rate  are  supplied  from the polar cap region
to the outer gap accelerator.  If the multipole fields in the south and north  poles could be an asymmetric configuration, the
asymmetric pair-creation process at the two polar caps could make
one outer gap, connecting to one pole,
less active than that connecting to the other pole. 

The high-energy emissions in the outer magnetosphere has been discussed with the
slot gap model (e.g. Arons 1983; Harding et al. 2008; Harding \& Kalapotharakos 2015),
the outer gap model (e.g. Cheng et al. 1986ab, Takata et al. 2011) and
the current sheet of the force-free or dissipative pulsar magnetosphere model
(e.g. Spitkovsky 2006; Kalapotharakos et al. 2014, 2017; Cerutti and Beloborodov 2017).
These models predict that  a larger viewing angle of  $\zeta\sim 90^{\circ}$
is preferred  to explain the double peak
structure of the GeV emission with the peak separation of $\sim 0.5$ phase for PSR~J2021+4026 (Takata et al. 2011;
Kalapotharakos et al. 2014). With a larger viewing angle,
on the  other hand, the emission from two polar caps that are equally heated up will make a double peak structure in the X-ray light curve, as demonstrated
in Figures~\ref{two-pole} and~\ref{two-pole-1} which
is inconsistent with the observation. This also motivates us to speculate that
one polar cap is less active.  

\begin{figure}
  \centering
  \includegraphics[scale=0.4]{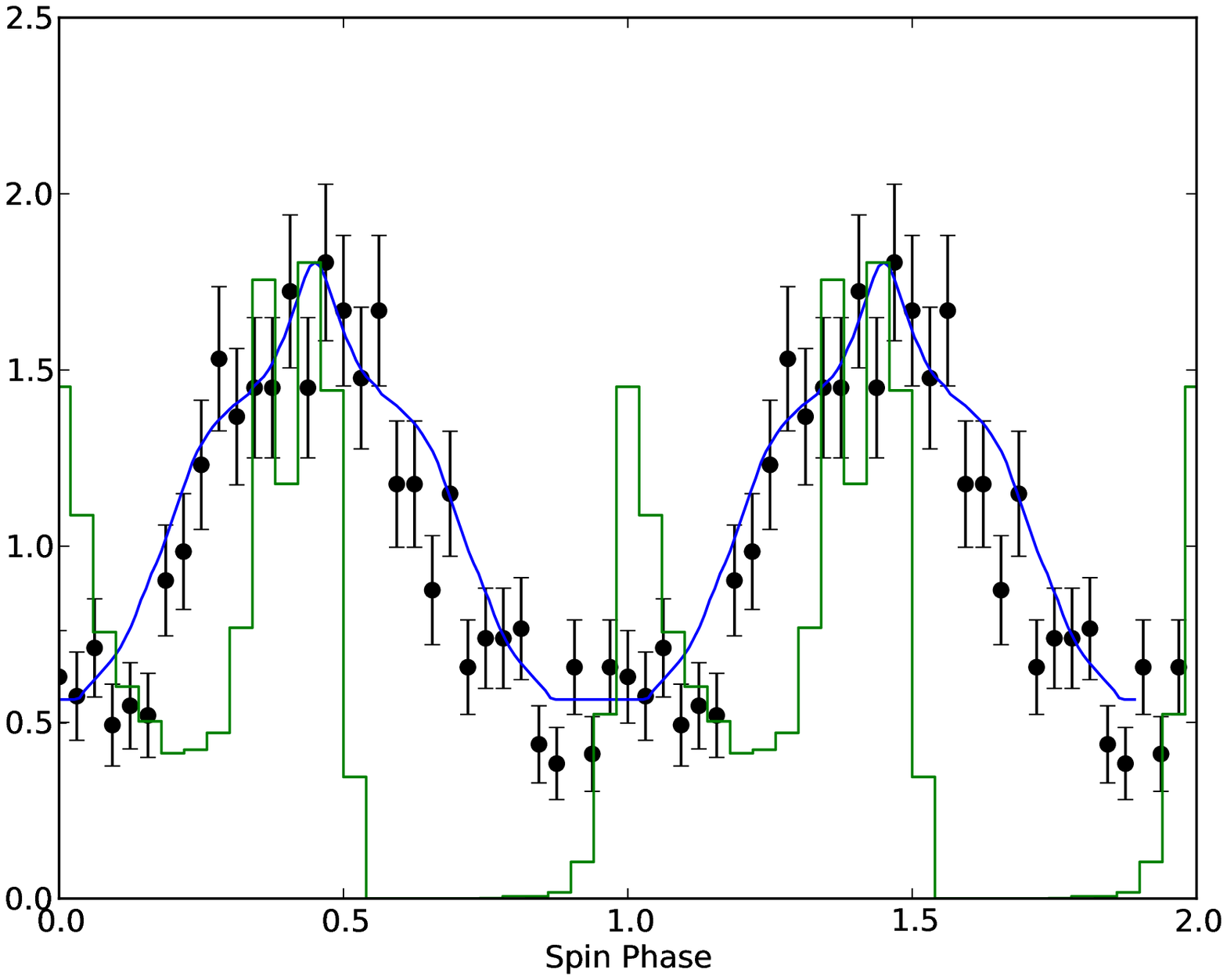}
  \includegraphics[scale=0.4]{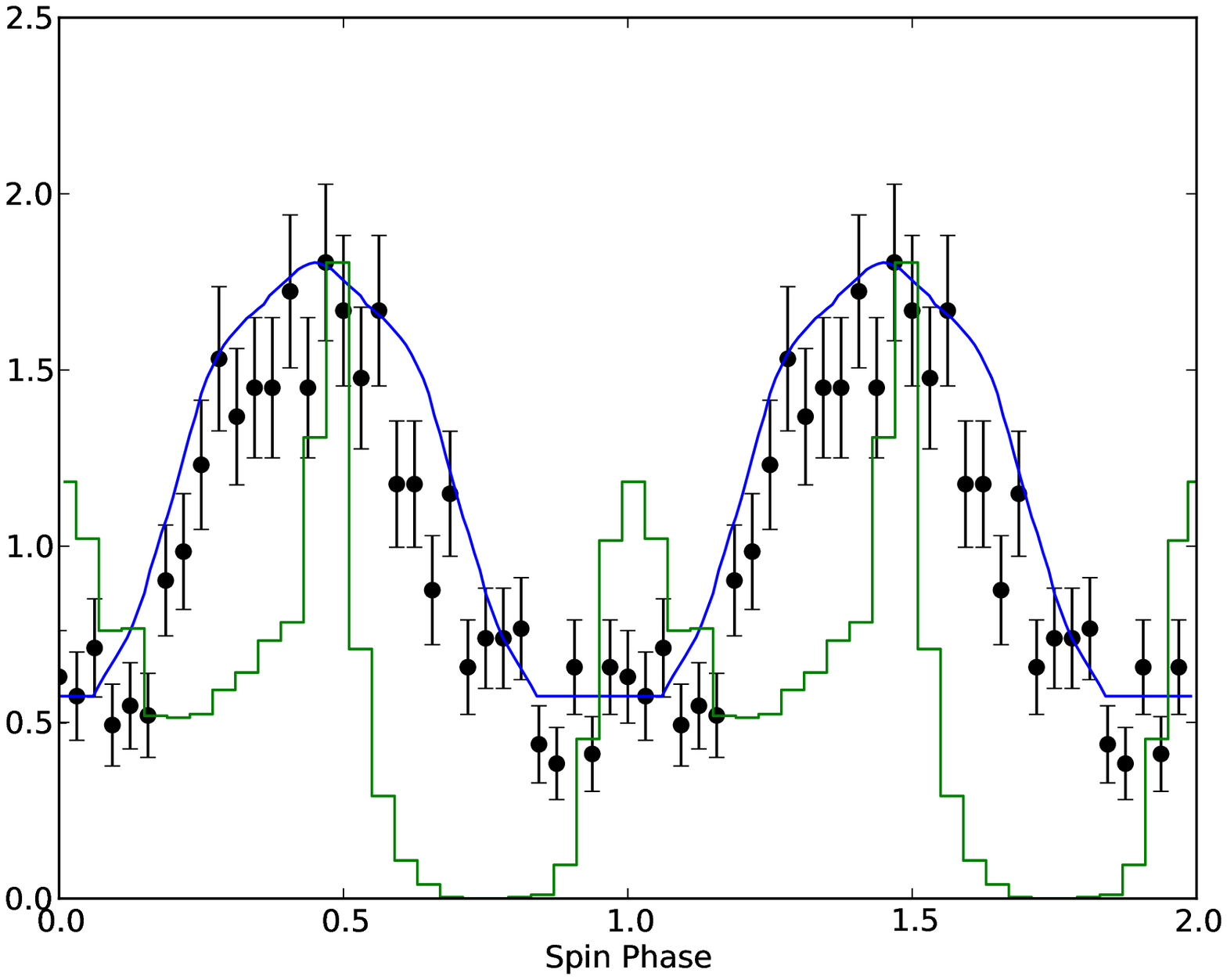}
  \caption{Model  X-ray (blue curve) and GeV (green histogram) light curves of PSR~J2021+4026. The emission geometry is the same as in Figure~\ref{geometry}, and
    the left (or right) panel assumes that only the outer gap connecting to north pole (or south pole) is active.
    The results are for the  inclination angle $\alpha=60^{\circ}$ and the
    viewing angle $\zeta=85^{\circ}$, which are chosen to explain both 
    observed phase-separation of the gamma-ray peaks and the
    pulsed fraction of the X-ray light curve.}
  \label{fit}
\end{figure}

Figure~\ref{fit} shows the model light curves of the X-ray from the heated polar cap and of the gamma-ray
from the outer gap accelerator by assuming that only one magnetic hemisphere is active, as illustrated by Figure~\ref{geometry};
we assume $\alpha=60^{\circ}$ and $\zeta=85^{\circ}$, and the outer gap
in the north hemisphere (left panel) or the south hemisphere (right panel) is active.  As in Figure~\ref{two-pole-1}, we
refer to the angular distribution of the local emission of the polar cap from Zavlin et al. (1995).  We find in the figure that
the single pole model can produce the single peak with a large pulsed fraction, and it is consistent with the observation.
To investigate the phase relation between
the gamma-ray pulse and the X-ray pulse, we apply  a simple outer gap model based on Wang et al. (2010),  and
consider both outwardly and inwardly propagating gamma-rays  from the outer gap. In Figure~\ref{fit}, we can see that the second peak of the gamma-ray light curve (green lines) is almost aligned with the X-ray peak, and the first peak is located at the minimum of the X-ray light curve.
These features are roughly consistent with the observed correlation in Figure~\ref{lightgx}. 
The current observation indicates that the X-ray peak shifts from the second peak of the gamma-ray light curve by the $\sim 0.14$~phase (section~\ref{timing}). 
This may indicate that realistic magnetic field geometry at the polar cap or at the outer gap may be different from the dipole field.

In summary, we have examined the properties of the X-ray emission from the variable gamma-ray pulsar, PSR~J2021+4026, in the low gamma-ray flux state and the post-relaxation state. 
The X-ray emission from both states are well described by a power law plus the blackbody radiation. 
The former is probably unpulsed and is originated from the pulsar wind nebula (Hui et al. 2015). 
The emission property of the blackbody component is consistent with the emission from the polar cap heated by the back-flow bombardment of the high-energy electrons or positrons that were accelerated in the magnetosphere.
We found no significant change in the X-ray emission properties at both states. The X-ray pulse profile could  be fitted better by the emission from one
pole rather than two poles, which suggests an asymmetric magnetosphere.

We express our appreciation to an anonymous referee for useful comments and suggestions.
We thank K.S. Cheng, C.-Y. Ng, C.Y. Hui, A.K. Kong, and P.H. Tam for the useful discussions.
W.H.H. and J.T. are supported by NSFC grants of the Chinese Government under 11573010, U1631103 and 11661161010.
LlL.C.C. is supported by the National Research Foundation of Korea  through grant 2016R1A5A1013277.


\newpage
\begin{thebibliography}{}
  \expandafter\ifx\csname natexlab\endcsname\relax\def\natexlab#1{#1}\fi
\bibitem[{Abdo} {et~al.}(2009a)]{abdo09a}
  {Abdo}, A.~A., {Ackermann}, M., {Ajello}, M., {Ampe}, J.,
  {Anderson}, B., {Atwood}, W.~B., {Axelsson}, M., {Bagagli}, R.,
  {Baldini}, L., {Ballet}, J., et al., 2009a, Astroparticle Physics, 32, 193
\bibitem[{Abdo} {et~al.}(2009b)]{abdo09b}
  {Abdo}, A.~A., {Ackermann}, M., {Ajello}, M., {Atwood}, W.~B., {Axelsson}, M. {et~al.},
  2009b, \apjs, 183, 46
\bibitem[{Abdo} {et~al.}(2013)]{abdo13}
  {Abdo}, A.~A., {Ajello}, M., {Allafort}, A., {Baldini}, L.,
    {Ballet}, J., {Barbiellini}, G., {Baring}, M.~G., {Bastieri}, D.,
    {Belfiore}, A., {Bellazzini}, R., et al., 2013, \apjs, 208, 17
  \bibitem[{Ackermann}{et~al}(2012)]{ack12}
    {Ackermann}, M., {Ajello}, M., {Albert}, A.,  {Allafort}, A.,
    {Atwood}, W.~B., {Axelsson}, M., {Baldini}, L., {Ballet}, J., et al, 2012, \apjs, 203, 4
    
  \bibitem[{Allafort}{et~al}(2013)]{allafort13}
   {Allafort}, A., {et al}. 2013, \apjl, 777, L2
 \bibitem[{Atons}(1983)]{at83}
   {Arons}, J., 1983, \apj, 266, 215
 \bibitem[{Atwood} {et~al.}(2009)]{at09}
   {Atwood}, W.~B., {Abdo}, A.~A., {Ackermann}, M., {Althouse}, W.,
   {Anderson}, B., {Axelsson}, M., {Baldini}, L., {Ballet}, J., 
   {Band}, D.~L., {Barbiellini}, G., et al., 2009, \apj, 697, 1071
   
 \bibitem[{Beloborodov}(2002)]{bel02}
   {Beloborodov}, A.~M., 2002, \apjl, 566, 85
\bibitem[{Bogdanov}(2016)]{bog16}
  {Bogdanov}, S., 2016, European Physical Journal A, 52, 37
\bibitem[{Bogdanov}{et~al}(2007)]{bog07}
  {Bogdanov}, S., {Rybicki}, G.~B. and {Grindlay}, J.~E., 2007, \apj, 670, 668
\bibitem[{Cerutti and Beloborodov}(2017)]{ce17}
  {Cerutti}, B. and {Beloborodov}, A.~M., 2017, \ssr, 207, 111
\bibitem[{Cheng}{et~al}(1986a)]{cheng86a}
  {Cheng}, K. S.,{ Ho}, C., {Ruderman}, M., 1986a,\apj , 300, 500
  \bibitem[{Cheng}{et~al}(1986b)]{cheng86b}
 { Cheng}, K. S.,{ Ho}, C., {Ruderman}, M., 1986b, \apj, 300, 522
\bibitem[{Cheng}{et~al}(1999)]{cheng99}
  {Cheng}, K. S., {Zhang}, L., 1999, \apj, 515, 337
\bibitem[{Daugherty and Harding}(1996)]{dau96}
  {Daugherty}, J.~K. and {Harding}, A.~K., 1996, \apj, 458, 278

\bibitem[{Gibson}{et~al}(1982)]{gib82}
Gibson, A. I., Harrison, A. B., Kirkman, I. W., et al. 1982, Natur, 296, 833
\bibitem[{Halpern} {et~al.}(1993)]{Hal93}
  {Halpern}, J.~P. and {Ruderman}, M., 1993, \apj, 415, 286
\bibitem[{Harding and  Kalapotharakos} (2015)]{ha15}
  {Harding}, A.~K. and {Kalapotharakos}, C., 2015, \apj, 811, 63
\bibitem[{Harding and Muslimov} (2011)]{ha11a}
  {Harding}, A.~K. and {Muslimov}, A.~G., 2011a, \apjl, 726, 10
\bibitem[{Harding and Muslimov} (2011)]{ha11b}
  {Harding}, A.~K. and {Muslimov}, A.~G., 2011b, \apj,743, 181
\bibitem[{Harding} {et~al.}(2008)]{Ha08}
  {Harding}, A.~K., {Stern}, J.~V., {Dyks}, J. and {Frackowiak}, M., 2008, \apj, 680, 1378
\bibitem[{Hirotani and Shibata} (2001)]{Hi01}
  {Hirotani}, K. and {Shibata}, S., 2001, \apj, 558, 216
\bibitem[{Ho and Lai} (2001)]{Ho01}
  {Ho}, W.~C.~G. and {Lai}, D., 2001, \mnras, 327, 1081
\bibitem[{Ho and Lai} (2003)]{Ho03}
    {Ho}, W.~C.~G. and {Lai}, D., 2003, \mnras, 338, 233
\bibitem[{Hui} {et~al.}(2015)]{Hui15}
  {Hui, }C. Y. ,{Seo}, K. A.  { Lin}, L. C. C.,  {Huang}, R. H. H. ,{Hu}, C. P.,  {Wu}, J. H. K., { Trepl}, L., {Takata}, J. , {et~al.},
  2015, \apj, 799, 1
\bibitem[{Kalapotharakos} {et~al.}(2017)]{ka17}
  {Kalapotharakos}, C., {Harding}, A.~K., {Kazanas}, D. and  {Brambilla}, G., 2017, \apj, 842, 80
\bibitem[{Kalapotharakos} {et~al.}(2014)]{ka14}
{Kalapotharakos}, C.,{Harding}, A.~K. and {Kazanas}, D., 2014, \apj, 793, 97 
\bibitem[{Lin} {et~al.}(2013)]{hui13}
  {Lin}, L. C. C.,{ Hu}, C. Y., {Hu}, C. P., {et al.}, 2013, \apjl, 770, L9
\bibitem[{Lin}{et~al}(2016)]{lin16}
  {Lin},L.C.C., 2016, JASS, 33, 147
\bibitem[{Leahy}(1987)]{lea87}
  Leahy, D. A. 1987, A\&A, 180, 275
\bibitem[{Mardia}(1972)]{ma72}
  Mardia, K. V. 1972, Statistics of Direction Data (New York: Academic)
\bibitem[{NG}{et~al}(2016)]{ng16}
 {Ng}, C. W., {Takata}, J. {Cheng}, K. S.,{et~at},  2016, \apj,  825, 18
\bibitem[{Pavlov}{et~al}(1994)]{pa94}
  {Pavlov}, G.~G., {Shibanov}, Y.~A., {Ventura}, J. and {Zavlin}, V.~E., 1994, \aap,  289, 837
\bibitem[{Pechenick}{et~al.}(1983)]{pe83}
{Pechenick}, K.~R., {Ftaclas}, C. and {Cohen}, J.~M., 1983, \apj, 274, 846
\bibitem[{Romani}(1995)]{ro95}
  {Romani}, R.~W. and {Yadigaroglu}, I.-A., 1995, \apj, 438, 314
  \bibitem[{Ruderman}(1988)]{ru88}
  {Ruderman}, M. and {Cheng}, K.S.,1988, \apj, 335, 306
 \bibitem[{Ruderman}(1991)]{ru91} 
   {Ruderman}, M.,1991, \apj, 366, 261
 \bibitem[{Spitkovsky}(2006)]{sp06}
   {Spitkovsky}, A., 2006, \apjl, 648, 51  
\bibitem[{Takata}{et~al}(2016)]{takata16}
  {Takata}, J., {Ng}, C.~W. and {Cheng}, K.~S., 2016, \mnras, 455, 4249
\bibitem[{Takata}{et~al}(2011)]{takata11}
  {Takata}, J., {Wang}, Y., {Cheng}, K. S., 2011, \mnras, 415, 1827
\bibitem[{Takata}{et~al}(2010)]{takata10}
  {Takata}, J. and {Wang}, Y. and {Cheng}, K.~S., 2010, \apj, 715, 13188
\bibitem[{Timokhin and Harding} (2015)]{tim15}
  {Timokhin}, A.~N. and {Harding}, A.~K., 2015, \apj, 810, 144
\bibitem[{Trepl} {et~al.}(2010)]{trel10}
 {Trepl}, L., {Hui}, C. Y., {Cheng}, K. S., {Takata}, J., {Wang}, Y.,{et~al},
  2010, \mnras, 405, 1339
\bibitem[{wang}{et~al}(2010)]{w010}
  {Wang}, Y. and {Takata}, J. and {Cheng}, K.~S., 2010, \apj, 720, 178
\bibitem[{weisskof}{et~al}(2011)]{weis11}
    {Weisskopf}, M.~C., {Romani}, R.~W., {Razzano}, M.,
    {Belfiore}, A., {Saz Parkinson}, P., {Ray}, P.~S.,
    {Kerr}, M., {Harding}, A., et al.  2011, \apj, 743, 74
  \bibitem[{Zane} {et~al.}(2000)]{zan00}
    {Zane}, S. and {Turolla}, R., 2006, \mnras, 366, 727
  \bibitem[{Zavlin} {et~al.}(1995)]{zav95}
    {Zavlin}, V.~E. and {Shibanov}, Y.~A. and {Pavlov}, G.~G., 1995, Astronomy Letters, 21, 149
    
 \bibitem[{Zhang}{et~al}(1997)]{zhang97}
 {Zhang}, L. \& {Cheng}, K. S. ,1997, \apj, 480, 370
\bibitem[{Zhao} {et~al.}(2017)]{zhao17}
  {Zhao}, J., {Ng}, C. W.,  {Lin}, L. C. C., {Takata}, J. , {Cai}, Y., {Hu}, C. P.,{et~al},
  2017, \apj, 842, 53

  
 
 
 
\end{thebibliography}
\end{document}